\documentclass{piparticle-final}
\usepackage[utf8]{inputenc}
\usepackage{graphicx}
\usepackage{amsmath}
\usepackage{amssymb}

\usepackage{cite} 
\usepackage{epstopdf}

\begin{document}
\newcommand{\be}{\begin{equation}}
\newcommand{\ee}{\end{equation}}
\newcommand{\ben}{\begin{eqnarray}}
\newcommand{\een}{\end{eqnarray}}
\newcommand{\nn}{\nonumber \\}
\newcommand{\ii}{\'{\i}}
\newcommand{\pp}{\prime}
\newcommand{\tr}{{\mathrm{Tr}}}
\newcommand{\nd}{{\noindent}}
\newcommand{\grad}{\hspace{-2mm}$\phantom{a}^{\circ}$}

\volume{7}               
\articlenumber{070006}   
\journalyear{2015}       
\editor{C. A. Condat, G. J. Sibona}   
\received{20 November 2014}     
\accepted{1 April 2015}   
\runningauthor{O. A. Rosso \itshape{et al.}}  
\doi{070006}         

\title{Noise versus chaos in a causal Fisher-Shannon plane}

\author{Osvaldo A. Rosso,\cite{inst1,inst2}\thanks{Email: oarosso@gmail.com}\hspace{0.5em}
	Felipe Olivares,\cite{inst3}\hspace{0.5em} Angelo Plastino\cite{inst4}
	}

\pipabstract{
We revisit the Fisher-Shannon representation plane ${\mathcal H} \times {\mathcal F}$, 
evaluated using the Bandt and Pompe recipe to assign a probability distribution to a
time series. Several stochastic dynamical (noises with $f^{-k}$,
$k \geq 0$, power spectrum) and chaotic processes (27 chaotic
maps) are analyzed so as to illustrate the approach. Our  main
achievement is uncovering the informational properties of the
planar location.
}

\maketitle

\blfootnote{
\begin{theaffiliation}{99}
	\institution{inst1} Insitituto Tecnol\'ogico de Buenos Aires, Av. Eduardo Madero 399, C1106ACD Ciudad Aut\'onoma de Buenos Aires, Argentina.
	\institution{inst2} Instituto de F\'isica, Universidade Federal de Alagoas, Macei\'o, Alagoas, Brazil.
	\institution{inst3} Departamento de F\'isica, Facultad de Ciencias Exactas, Universidad Nacional de La Plata, La Plata, Argentina.
	\institution{inst4} Instituto de F\'isica, IFLP-CCT, Universidad Nacional de La Plata, La Plata, Argentina.
\end{theaffiliation}
}


%

\section{Introduction}

Temporal sequences of measurements (or observations), that is,
time-series (TS), are the basic elements  for investigating
natural phenomena. From TS, one should judiciously extract
information on dynamical systems. Those TS  arising from chaotic
systems share with those generated by stochastic processes several
properties that make them very similar: (1) a wide-band power
spectrum (PS), (2) a delta-like autocorrelation function, (3) 
irregular behavior of the measured signals, etc. Now, irregular
and apparently unpredictable behavior is often observed in natural
TS, which makes interesting the  establishment of  whether the
underlying dynamical process is of either deterministic or
stochastic character in order to {\it i)\/} model the associated
phenomenon and {\it ii)\/} determine which are the relevant
quantifiers.

Chaotic systems display  ``sensitivity to initial conditions'' and
lead to non-periodic motion (chaotic time series). Long-term
unpredictability arises despite the deterministic character of the
 trajectories (two
neighboring points in the phase space move away exponentially
rapidly). Let ${\mathbf x}_1(t)$ and ${\mathbf x}_2(t)$ be two
such points,  located within a ball of radius $R$ at time $t$.
Further, assume that these two points cannot be resolved within
the ball due to poor instrumental resolution. At some later time
$t'$, the distance between the points will typically grow to
$|{\mathbf x}_1(t')-{\mathbf x}_2(t')| \approx |{\mathbf
x}_1(t)-{\mathbf x}_2(t)|~\exp( \lambda~|t'-t|)$, with $\lambda >
0$ for a chaotic dynamics, $\lambda$ the largest Lyapunov
exponent. When this distance at time $t'$ exceeds $R$, the points
become experimentally distinguishable. This  implies that
instability reveals some information about the phase space
population that was not available at earlier times
\cite{Abarbanel1996}. One can then think of chaos as an {\sl
information source\/.} The associated rate of generated
information can be cast in  precise fashion via the
Kolmogorov-Sinai's entropy \cite{Kolmogorov1958,Sinai1959}.

 One question often emerges:   is the system chaotic (low-dimensional
deterministic) or stochastic?  If one is able to show that the
system is dominated by low-dimensional deterministic chaos, then
only few (nonlinear and collective) modes are required to describe
the pertinent dynamics \cite{Osborne1989}. If not, then the
complex behavior could be modeled by a system dominated by a very
large number of excited modes which are in general better
described by stochastic or statistical approaches.

Several methodologies for evaluation of Lyapunov exponents and
Kol\-mogorov-Sinai entropies for time-series' analysis have been
proposed (see Ref. \cite{Kantz2002}), but their applicability
involves taking into account constraints (stationarity, time
series length, parameters values election for the methodology,
etc.) which in general make the ensuing results {\sl
non-conclusive}. Thus, one wishes for new tools able to
distinguish chaos (determinism) from noise (stochastic) and this
leads to our present interest in the computation of quantifiers
based on Information Theory, for instance, ``entropy'', ``statistical
complexity'', ``Fisher information'', etc.

These quantifiers can be used to detect determinism in time series
\cite{Rosso2007,Rosso2011,Rosso2012,Rosso2013,Olivares2012A,Olivares2012B}.
Different Information Theory based measures (normalized Shannon
entropy, statistical complexity, Fisher information) allow for a
better distinction between deterministic chaotic and stochastic
dynamics whenever ``causal'' information is incorporated via the
Bandt and Pompe's (BP) methodology \cite{Bandt2002}. For a review
of BP's methodology and its applications to physics, biomedical
and econophysic signals, see \cite{Zanin2012}.

Here we revisit, for the purposes previously detailed, the so-called
causality Fisher--Shannon entropy plane, ${\mathcal H} \times
{\mathcal F}$ \cite{Vignat2003}, which allows to quantify the
global versus local characteristic of the
 time series generated by the dynamical process under study. The
two functionals ${\mathcal H}$ and ${\mathcal F}$ are evaluated
using the Bandt and Pompe permutation approach.  Several
stochastic dynamics (noises with $f^{-k}$, $k \geq 0$, power
spectrum) and chaotic processes (27 chaotic maps) are analyzed so
as to illustrate the methodology. We will encounter that
significant information is provided by the planar location.

\section{Shannon entropy and Fisher information measure}
\label{sec:Shannon-Fisher}

Given a continuous probability distribution function (PDF) $f(x)$
with $x \in \Delta \subset {\mathbb R}$ and $\int_{\Delta} f(x)~dx = 1$,
its associated {\it Shannon Entropy\/} $S$  \cite{Shannon1949} is
\begin{equation}
\label{shannon}
S[f]~=~-\int_{\Delta}~f~\ln(f)~dx \ ,
\end{equation}
a measure of ``global character'' 
that is not too sensitive to strong changes in the distribution taking place on a
small-sized region. Such is not the case with {\it Fisher's
Information Measure\/} (FIM) $\mathcal F$ \cite{Fisher1922,Frieden2004},
which constitutes a measure of the gradient content of the
distribution $f(x)$, thus being quite sensitive even to tiny
localized perturbations. It reads
\begin{align}
\label{fisher}
{\mathcal F}[f] &= \int_{\Delta}~ { {1} \over {f(x)} } \left[ { {df(x)} \over {dx} }\right]^2 ~dx \notag \\
 &= 4 \int_{\Delta}~\left[ { {d \psi(x)} \over {dx} }\right]^2.
\end{align}
FIM can be variously interpreted as a measure
of the ability to estimate a parameter, as the amount of
information that can be extracted from a set of measurements, and
also as a measure of the state of disorder of a system  or
phenomenon \cite{Frieden2004}.
In the previous definition of FIM (Eq. (\ref{fisher})), the
division by $f(x)$ is not convenient  if $f(x) \rightarrow 0$ at
certain $x-$values. We avoid this if we
work with real probability amplitudes $f(x)= \psi^{2}(x)$
\cite{Fisher1922,Frieden2004},
which is a  simpler form (no divisors) and
shows that $\mathcal F$ simply measures the gradient content in $\psi(x)$.
The gradient operator significantly influences the
contribution of minute local $f-$variations to FIM's value.
Accordingly, this quantifier is called a ``local'' one
\cite{Frieden2004}.

Let now $P=\{p_i;~i=1,\cdots, N\}$  be a  discrete probability
distribution, with $N$ the number of possible states of the system
under study. The concomitant  problem of information-loss due to
discretization has been thoroughly studied
and, in particular, it entails the loss of FIM's shift-invariance,
which is of no importance for our present purposes \cite{Olivares2012A,Olivares2012B}.
In the discrete case,  we define a ``normalized'' Shannon entropy as
\begin{equation}
\label{shannon-disc}
{\mathcal H}[P]~=~ \frac{S[P]}{S_{max}} ~=~ \frac{1}{S_{max}}\left\{-\sum_{i=1}^{N}~p_i~\ln( p_i) \right\},
\end{equation}
where the denominator  $S_{max} = S[P_e] = \ln N$ is that attained
by a uniform probability distribution $P_e = \{p_i =1/N,~ \forall i = 1, \cdots, N\}$.
For the FIM, we take the expression in term of  real probability amplitudes
as starting point, then a discrete normalized FIM convenient for our present purposes
is given by
\begin{equation}
\label{Fisher-disc}
{\mathcal F}[P]~=~F_0~\sum_{i=1}^{N-1}~[(p_{i+1})^{1/2} - (p_{i})^{1/2}]^2 \ .
\end{equation}
It has been extensively discussed that this discretization is the best behaved in a 
discrete environment \cite{Jesus2009}. Here, the normalization constant $F_0$ reads
\begin{equation}
\label{F0}
F_0~=~\left\{
       \begin{array}{cl}
                    1       &\qquad \mbox{\footnotesize if $p_{i^*} = 1$ for $i^* = 1$ or} \\
                            &\qquad \mbox{\footnotesize $i^* = N$ and $p_{i}  = 0 ~\forall  i \neq i^*$} \\
                    1/2     &\qquad \mbox{otherwise}.
       \end{array}
\right.
\end{equation}
If our system lies in a very ordered state, which occurs when almost all the $p_{i}$ -- 
values are zeros, we have a normalized Shannon entropy ${\mathcal H} \sim 0$ 
and a normalized Fisher's Information Measure ${\mathcal F} \sim 1$.
On the other hand, when the system under study is represented by  a very disordered state, 
that is when all the $p_{i}$ -- values oscillate
around the same value, we obtain ${\mathcal H} \sim 1$ while ${\mathcal F} \sim 0$.
One can state that the general FIM-behavior of the present discrete version (Eq. (\ref{Fisher-disc})),  
is opposite to that of the Shannon entropy, 
except for periodic motions \cite{Olivares2012A,Olivares2012B}.
The local sensitivity of FIM for  discrete-PDFs is reflected in the fact that the specific
``$i-$ordering'' of the discrete values $p_{i}$ must be seriously taken into account in
evaluating the sum in Eq. (\ref{Fisher-disc}).
This point was extensively discussed by us in previous works \cite{Olivares2012A,Olivares2012B}.
The summands can be regarded as a kind of ``distance'' between  two contiguous probabilities.
Thus, a different ordering of the pertinent summands would lead to a different FIM-value,
hereby its local nature.
In the present work, we follow the lexicographic order described by Lehmer \cite{Lehmer} in the
generation of Bandt-Pompe PDF.

\section{Description of our chaotic and stochastic systems}
\label{sec:Systems}

Here we study both chaotic and stochastic systems,  selected  as
illustrative examples of different classes of signals, namely,
{\it (a)\/} 27 chaotic dynamic maps \cite{Rosso2013,Sprott2003} and
{\it (b)\/} truly stochastic processes,  noises with $f^{-k}$ power spectrum \cite{Rosso2013}.

\subsection{Chaotic maps}
\label{sec:Chaotic-maps}

In the present work, we consider 27 chaotic maps described by J. C.
Sprott in the appendix of his book \cite{Sprott2003}.
These chaotic maps are grouped  as
\begin{itemize}
\item  [ {\it a)\/} ]{\it  Noninvertible maps:\/}
(1) Logistic map; 
(2) Sine map; 
(3) Tent map; 
(4) Linear congruential generator; 
(5) Cubic map; 
(6) Ricker's population model; 
(7) Gauss map; 
(8) Cusp map; 
(9) Pinchers map; 
(10) Spence map; 
(11) Sine-circle map; 
\item [ {\it b)\/} ] {\it Dissipative maps:\/}
(12) H\'enon map; 
(13) Lozi map; 
(14) Delayed logistic map; 
(15) Tinkerbell map; 
(16) Burgers' map; 
(17) Holmes cubic map; 
(18) Dissipative standard map; 
(19) Ikeda map; 
(20) Sinai map; 
(21) Discrete predator-prey map, 
\item [ {\it c)\/} ] {\it Conservative maps:\/}
(22) Chirikov standard map; 
(23) H\'enon area-preserving quadratic map; 
(24) Arnold's cat map; 
(25) Gingerbreadman map; 
(26) Chaotic web map; 
(27) Lorenz three-dimensional chaotic map; 
\end{itemize}
Even when the present list of chaotic maps is not exhaustive, it could
be taken as representative of common chaotic systems \cite{Sprott2003}.

\subsection{Noises with $f^{-k}$ power spectrum}
\label{sec:Noises}

The corresponding time series  are generated as follows
\cite{Hilda2012}: 1) Using the Mersenne twister generator
\cite{mersenne1998} through the $\textsc{Matlab}^\copyright$ {\sl
RAND\/} function we generate pseudo random numbers $y^0_i$ in the
interval $(-0.5,0.5)$ with an {\it (a)\/} almost flat power
spectra (PS), {\it (b)\/} uniform PDF, and {\it (c)\/} zero mean
value. 2) Then, the Fast Fourier Transform (FFT) ${y^1_i}$ is
first obtained and then multiplied by $f^{-k/2}$, yielding
${y^2_i}$; 3) Now, ${y^2_i}$ is symmetrized so as to obtain a real
function. The pertinent inverse FFT is now at our disposal, after
discarding the small imaginary components produced by the
numerical approximations.
The resulting time series $\eta^{(k)}$  exhibits the desired power spectra and,
by construction,  is representative of non-Gaussian noises.

\section{Results and discussion}
\label{sec:Results}

In all chaotic maps, we took (see section \ref{sec:Chaotic-maps}) the same
initial conditions and the parameter-values detailed by Sprott.
The corresponding initial values are given in the basin of
attraction or near the attractor for the dissipative systems, or
in the chaotic sea for the conservative systems \cite{Sprott2003}.
For each map's TS,  we discarded the first $10^5$ iterations and,
after that, $N = 10^7$ iterations-data were generated. 

Stochastic dynamics represented by time series of noises with $f^{-k}$ power
spectrum ($0 \leq k \leq 3.5$ and $\Delta k = 0.25$) were
considered. For each value of $k$, ten series with different seeds
and total length $N=10^6$ data were generated (see section
\ref{sec:Noises}), and their corresponding average values were
reported for uncorrelated ($k=0$) and correlated ($k>0$) noises.

\begin{figure}[tbp] 
\begin{center}
\includegraphics[width = 0.98\columnwidth]{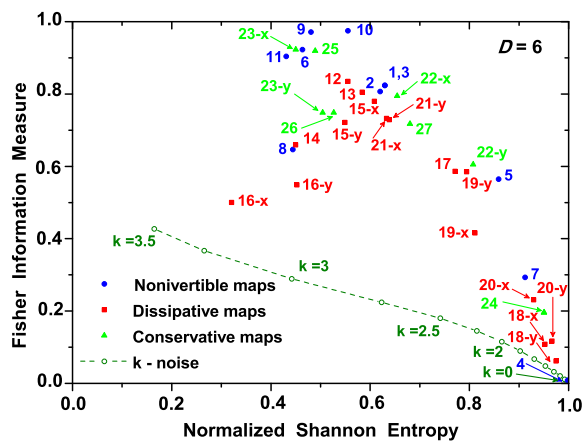}
\caption{
Localization in the causality Fisher-Shannon plane
of the 27 chaotic maps considered in the present work. The
Bandt-Pompe PDF was evaluated following the lexicographic order \cite{Lehmer} and
considering $D = 6$ (pattern-length),
$\tau = 1$ (time lag) and  time series length $N = 10^7$ data
(initial conditions given by Sprott \cite{Sprott2003}).
The  inside numbers represent the corresponding chaotic map
enumerated at the beginning of section \ref{sec:Chaotic-maps}.
The letters ``X'' and ``Y'' represent the time series coordinates maps
for which their planar representation is clearly distinguishable.
The open circle-dash line represents the planar localization
(average values over ten realizations with different seeds)
for the stochastic process: noises with $f^{-k}$ power spectrum.}
\label{fig:HxF-27-mapas}
\end{center}
\end{figure}

The BP-PDF was evaluated for each TS of $N$ data, stochastic and
chaotic, following the lexicographic pattern-order proposed by
Lehmer \cite{Lehmer}, with pattern-lengths $D = 6$ and time lag
$\tau = 1$. Their corresponding localization in the causality
Fisher-Shannon plane are shown in Fig. \ref{fig:HxF-27-mapas}. One
can use any of these TS for  evaluating the dynamical system's
invariants (like correlation dimension, Lyapunov exponents, etc.),
by appealing to a time lag reconstruction \cite{Sprott2003}.  Here
we analyzed TS generated by each one of chaotic maps' coordinates
when the corresponding map is bi- or multi-dimensional.
Due to the fact that the BP-PDF is not a dynamical invariant
(neither are other quantifiers derived by Information Theory), some
variation could be expected in the quantifiers' values computed
with this PDF, whenever one or other of the TS generated by
these multidimensional coordinate systems.

From Fig. \ref{fig:HxF-27-mapas}, we clearly see that the chaotic maps under study are localized
mainly at entropic region lying between $0.35$ and $0.9$, and reach FIM values from $0.4$ to almost $1$.
A second group of chaotic maps, constituted by:
the Gauss map (7),
linear congruential generator (4),
dissipative standard map (18),
Sinai map (20) and
Arnold's cat map,
is localized near the right-lower corner of the ${\mathcal H} \times {\mathcal F}$ plane, that is in the
range $0.95 \leq {\mathcal H} \leq 1.0$ and $0 \leq {\mathcal F} \leq 0.3$.
Their localization could be understood if one takes into account that when a 2D-graphical representation
of them (i.e., a graph $X_n \times X_{n+1}$ for one dimensional maps, or $X_n \times Y_n$ for two dimensional maps)
it tends to fulfill the space, resembling the behavior of stochastic dynamics. However, they are chaotic and
present a clear structure when the dynamics are represented in higher dimensional plane.

Noises with $f^{-k}$ power spectrum (with $0 \leq k \leq 5$)
exhibit a wide range of entropic values ($0.1 \leq {\mathcal H}\leq 1$) and
FIM values lying between $0 \leq {\mathcal F} \leq 0.5$. A smooth transition
in the planar location is observed in the passage from
uncorrelated noise ($k=0$ with ${\mathcal H} \sim 1$ and ${\mathcal F} \sim 0$) to
correlated one ($k>0$). The correlation degree grows as the $k$
value increases. From Fig.~\ref{fig:HxF-27-mapas} we gather that,
for  stochastic time series with increasing correlation-degree,
the associated entropic values ${\mathcal H}$ decrease, while Fisher's values
${\mathcal F}$ increase.
Taking into account that other stochastic processes, like fBm and fGn (not shown),
present a quite close behavior to the $k$-noise analyzed here (see Ref. \cite{Olivares2012B}), we can think that the open circle-dash
line represents a division of the plane; above this line all the chaotic maps are localized.
It is also interesting to note that, qualitatively, the same results are obtained when the evaluations where
made with pattern length $D=4$ and $D=5$, as well as, different Fisher information measure discretization
are used.

Summing up, we have presented an extensive series of numerical
simulations/computations and have contrasted the characterizations
of deterministic chaotic  and noisy-stochastic dynamics, as
represented by time series of finite length. {\it Surprisingly
enough, one just has to look at the different planar locations of
our two dynamical regimes}. The planar location is able to tell us
whether we deal with chaotic or stochastic time series.

\begin{acknowledgements}
O. A. Rosso and A. Plastino were supported by Consejo Nacional
de Investigaciones Cient\'{\i}ficas y T\'ecnicas (CONICET),
Argentina.
O. A. Rosso acknowledges support as a FAPEAL fellow, Brazil.
F. Olivares is supported by Departamento de F\'isica, Facultad de Ciencias Exactas, 
Universidad Nacional de La Plata, Argentina.
\end{acknowledgements}

\end{document}